\documentclass[fleqn,11pt]{article}

\usepackage{amsmath,amssymb}
\usepackage{amsfonts,amssymb,cite}
\usepackage{graphicx}

\def\noi{\noindent}

\newcommand{\Title}[1]{\noi {{\Large\bf #1}}\\[1ex]}

\newcommand{\Author}[2]{\noi{\bf #1}\\[2ex]\noi{\normalsize\it #2}\\}

\newcommand{\Abstract}[1]{\vskip 2mm \begin{center}
        \parbox{16.4cm}{\small\noi #1} \end{center}\medskip}
\newcommand{\foom}[1]{\protect\footnotemark[#1]}

\def\nqq{\hspace*{-2em}}

\def\Jl#1#2{#1 {\bf #2},\ }

\def\ApJ#1 {\Jl{Astroph. J.}{#1}}
\def\CQG#1 {\Jl{Class. Quantum Grav.}{#1}}
\def\DAN#1 {\Jl{Dokl. AN SSSR}{#1}}
\def\GC#1 {\Jl{Grav. Cosmol.}{#1}}
\def\GRG#1 {\Jl{Gen. Rel. Grav.}{#1}}
\def\JETF#1 {\Jl{Zh. Eksp. Teor. Fiz.}{#1}}
\def\JETP#1 {\Jl{Sov. Phys. JETP}{#1}}
\def\JHEP#1 {\Jl{JHEP}{#1}}
\def\JMP#1 {\Jl{J. Math. Phys.}{#1}}
\def\NPB#1 {\Jl{Nucl. Phys. B}{#1}}
\def\NP#1 {\Jl{Nucl. Phys.}{#1}}
\def\PLA#1 {\Jl{Phys. Lett. A}{#1}}
\def\PLB#1 {\Jl{Phys. Lett. B}{#1}}
\def\PRD#1 {\Jl{Phys. Rev. D}{#1}}
\def\PRL#1 {\Jl{Phys. Rev. Lett.}{#1}}

\def\lal{&&\nqq {}}

\def\beq{\begin{equation}}
\def\eeq{\end{equation}}
\def\bear{\begin{eqnarray}}
\def\bearr{\begin{eqnarray} \lal}
\def\ear{\end{eqnarray}}
\def\earn{\nonumber \end{eqnarray}}

\newcommand{\Fig}[3]{%
\begin{center}
\parbox{8cm}{%
\refstepcounter{figure}\includegraphics[width=8cm,height=#2cm]{#1} \noindent Figure \thefigure:\quad
#3}\end{center}}
\begin{document}
\thispagestyle{empty}
\twocolumn[


\Title{A Complete Model of Cosmological Evolution of Scalar Field with Higgs Potential and Euclidian Cycles\foom 1}

\Author{Yu.G. Ignat'ev, D.Yu. Ignatyev}
    {Institute of Physics, Kazan Federal University, Kremlyovskaya str., 18, Kazan, 420008, Russia}


\Abstract
 {The revision of the Author's results with respect to possibility of existence of the so-called Euclidian cycles in cosmological evolution of a system of Higgs scalar fields has been performed. The assumption of non-negativity of the Universe's extension velocity, which contradicts in certain cases to complete system of Einstein equations, has been removed. It has been shown that in cases when effective energy of the system tends to zero, a smooth transition of the model to the range of negative values of the extension velocity occurs, i.e. it occurs the transition to collapse stage rather than winding of the phase trajectories on the boundary of prohibited area. This process has been researched with a help of numerical simulation methods for the model based on classical scalar Higgs field.
}
\bigskip

] 

%
\section{Introduction}
The hypothesis assuming the possibility of existence of so-called \emph{Euclidian cycles} in cosmological models based on both classical and phantom scalar fields, was formulated by the Author in the series of recent works \cite{IgnKokh1}, \cite{IgnKokh2,{IgnKokh2a}}, \cite{IgnAgaf1}. The essence of the Euclidian cycles is concluded in tending of cosmological models at certain parameters of field model to the state with null effective energy. In this case, space-time becomes pseudo-euclidian although scalar fields are non-zero and perform free nonlinear oscillations. This assumption was formulated on the basis of analysis of qualitative and numerical investigations of cosmological models with scalar fields (see e.g. \cite{Ignat_Agaf_Mih_Dima15_3_AST},  \cite{Ignat_Sasha_G&G} along with the works cited above). In this paper we do not provide a review of papers devoted to cosmological models with scalar fields of other authors unless they are related directly to the essence of the investigation. A quite detailed review of such kind can be found in the cited above works \cite{IgnKokh1}, \cite{IgnAgaf1}. Let us however notice a series of works authored by I.A. Aref'eva, I.V. Volovich, N.V. Bula\-tov, R.V. Corbachev and S.Yu. Vernov \cite{Aref-Vol}, \cite{Aref-Vern}, which can be considered close to the matter of the current paper which were not known to the Author before. The cited papers consider states with null energy in the context of non-minimal models with negative cosmological constant. Let us notice the research by S.Yu. Vernov \cite{Vernov}, providing,  particularly, a quite satisfactory substantiation of cosmological models with phantom fields.

The author (Yu. Ignat'ev) was advised in a private conversation by Sergey Vernov that the system of dynamic equations describing cosmological evolution that was considered in papers \cite{IgnKokh1}, \cite{IgnKokh2,{IgnKokh2a}}, \cite{IgnAgaf1} is arguably not equivalent to the initial system of Einstein equations. More specifically, the limitation of the Hubble constant by non-negative values can contradict to the initial system of Einstein equations. The given work is devoted to the clarification of this circumstance and associated necessary correction of the results.

\section{Base Relations of the Model}
\subsection{Field Equations}
Let us consider a self-consistent system of Einstein equations of classical scalar field $\Phi$ with Higgs potential as a field model. The following Lagrangian function corresponds to scalar field $\Phi$
\begin{equation}\label{Ls}
\mathrm{L}_s=\frac{1}{8\pi}\biggl(\frac{1}{2}g^{ik}\Phi_{,i}\Phi_{,k}-V(\Phi)\biggr),
\end{equation}
where $V(\Phi)$ is a potential energy of the scalar field which for the Higgs potential takes the form:
\begin{equation}\label{Higgs}
V(\Phi )=-\frac{\alpha }{4} \left(\Phi ^{2} -\frac{e m^{2} }{\alpha } \right)^{2} ,
\end{equation}
$\alpha$ is a constant of self-action, $m$ is a mass of scalar bosons, $e=\pm 1$ is an indicator \footnote{Normally the case $e=+1$ is considered however we will not limit ourselves with this since alternative model is also of interest at $\alpha<0$.}.

Let us with the help of standard procedure get from the Lagrangian function \eqref{Ls} the equation of the scalar field:
\begin{equation}\label{Eq_s}
\Box\Phi+V'_\Phi=0.
\end{equation}
Then %
\begin{equation}\label{T_{iks}}
T^i_k=\frac{1}{8\pi}\bigl(\Phi^{,i}\Phi_{,k}-\frac{1}{2}\delta^i_k\Phi_{,j}\Phi^{,j}+\delta^i_k m^2V(\Phi)\bigr)
\end{equation}
is a tensor of the scalar field's energy - momentum. Let us further omit the constant term in the Higgs potential \eqref{Higgs} as it leads to simple redefinition of the cosmological term $\lambda$.

Corresponding Einstein equations of the considered system have the form\footnote{It is $G=\hbar=c=1$ everywhere in this paper, the signature of metrics is $(-,-,-,+)$, Ricci tensor is obtained by means of convolution of first
and third indices.}:
\begin{eqnarray}\label{Eq_Einst}
G^i_k\equiv R^i_k-\frac{1}{2}R\delta^i_k =8\pi T^i_k+\lambda\delta^i_k.
\end{eqnarray}
The equations \eqref{Eq_s} and (\ref{Eq_Einst}) are the base model of the cosmological scenario.
\subsection{The Dynamic Equations of the Space Flat Friedmann Model}
In the case of space flat Friedmann Universe
\begin{equation}\label{Freed}
ds^2_0=dt^2-a^2(t)(dx^2+dy^2+dz^2)
\end{equation}
the complete system of dynamic equations with respect to scale factor $a(\eta)$ and scalar potential $\Phi(\eta)$ takes the form:
\begin{eqnarray}\label{Eq_Phi0}
\ddot{\Phi}+3\frac{\dot{a}}{a}\dot{\Phi}+m^2 \Phi-\alpha\Phi^3=0;\\
\label{EqEinst0_4}
3\frac{\dot{a}^2}{a^2}-\frac{\dot{\Phi}^2}{2}-\frac{e m^2\Phi^2}{2}+\frac{\alpha\Phi^4}{4} - \lambda=0;\\
\label{EqEinst0_1}
2\frac{\ddot{a}}{a}+\frac{\dot{a}^2}{a^2}+\frac{\dot{\Phi^2}}{2}-\frac{e m^2\Phi^2}{2}+\frac{\alpha\Phi^4}{4}-\lambda=0.
\end{eqnarray}
First and foremost, let us make the following notice. Differentiating equation \eqref{EqEinst0_4} over time variable, we find the equality:
\begin{eqnarray}
\frac{3\dot{a}}{a}\biggl(2\frac{\ddot{a}}{a}-2\frac{\dot{a}\ \!^2}{a^2}+\dot{\Phi}^2\biggr)-\nonumber\\
\dot{\Phi}\biggl[\ddot{\Phi}+3\frac{\dot{a}}{a}\dot{\Phi}+e m^2 \Phi-\alpha\Phi^3\biggr]=0.\nonumber
\end{eqnarray}
As a consequence of the field equation \eqref{Eq_Phi0} we find from here
\begin{equation}\label{a'(11-44)}
\frac{\dot{a}}{a}\biggl(\frac{\ddot{a}}{a}-\frac{\dot{a}\ \!^2}{a^2}+\frac{1}{2}\dot{\Phi}^2\biggr)=0.
\end{equation}
Substituting the expression for $(-3\dot{a}^2/a^2)$ from the Einstein equation \eqref{EqEinst0_4} in \eqref{a'(11-44)} at  $\dot{a}\not\equiv0$,
we obtain the Einstin equation \eqref{EqEinst0_1}. Thus, the Einstein equation for the components $^\alpha_\alpha$ \eqref{EqEinst0_1} at $\dot{a}\not\equiv0$ is an algebraic - differential consequence of the field equation \eqref{Eq_Phi0} and Einstein equation for the component $^4_4$. We can consider the system of equations \eqref{Eq_Phi0} and difference of equations \eqref{EqEinst0_1} -- \eqref{EqEinst0_4} as a base dynamic system of space - flat cosmological model:
\begin{equation}\label{(11-44}
\frac{\ddot{a}}{a}-\frac{\dot{a}\ \!^2}{a^2}+\frac{1}{2}\dot{\Phi}^2=0.
\end{equation}
Introducing  \emph{the Hubble} ``\emph{constant}''
 \begin{equation}\label{H}
 H(t)=\frac{\dot{a}}{a},
 \end{equation}
 let us re-write the system of equations in the following form:
\begin{eqnarray}
\label{Eq_Phi1}
\ddot{\Phi}=&\displaystyle -3H\dot{\Phi}- e m^2 \Phi+\alpha\Phi^3;\\
\label{dH/dt}
\dot{H}=&\displaystyle -\frac{1}{2}\dot{\Phi}^2.\hspace{2.7cm}
\end{eqnarray}
However, here we should take into account the Einstein equation for the component $^4_4$ \eqref{EqEinst0_4} as it is an equation of the 1 order with respect to scalar potential and scale factor and limits the arbitrariness of the system's \eqref{Eq_Phi1} -- \eqref{dH/dt} solutions, practically being the first \emph{fixed} integral of this system.

Let us now introduce a value which will be of use in the future -- the effective energy of the system, $\mathcal{E}$:
\begin{equation}\label{E}
\mathcal{E}=\frac{\dot{\Phi}^2}{2}+\frac{e m^2\Phi^2}{2}-\frac{\alpha\Phi^4}{4}+\lambda,
\end{equation}
by means of which equations \eqref{EqEinst0_4} and \eqref{Eq_Phi1}, correspondingly, can be given the following compact form:
\begin{eqnarray}
\label{EqEinst0_4_E}
3H^2-\mathcal{E}=0,\\
\label{Eq_Phi2}
\dot{\mathcal{E}}+3H\dot{\Phi}^2=0;
\end{eqnarray}
Substituting the expression for $\dot{\Phi}^2$ from \eqref{dH/dt} into \eqref{Eq_Phi2}, we get the total energy's conservation law
\beq\label{d(3H^2-E)=0}
\dot{E}\equiv \frac{d}{dt}(3H^2-\mathcal{E})=0,
\eeq
with Einstein equation  \eqref{EqEinst0_4} for the space with null 3-dimensional curvature being a particular integral $E=0$ of it. 
Due to the non-negativity of $H^2$, an important property follows from \eqref{EqEinst0_4_E}:
\begin{equation}\label{E>=0}
\mathcal{E}\geqslant 0.
\end{equation}
Since, as was noted earlier, the Einstein equation \eqref{EqEinst0_4} and, in new notation, equation \eqref{EqEinst0_4_E} is a first particular integral of the field equations \eqref{Eq_Phi0} and Einstein equation \eqref{EqEinst0_1}, the relation \eqref{E>=0} is a necessary integral condition for the dynamic system \eqref{Eq_Phi0} -- \eqref{EqEinst0_1}.

Next, to get rid off the mentioned arbitrariness, let us consider the system of field equations \eqref{Eq_Phi0} and \eqref{EqEinst0_1} as a base system of equations, however let us use the definition of the Hubble constant \eqref{H} and expression for $\dot{\Phi}^2$ from \eqref{EqEinst0_4} in the equation \eqref{EqEinst0_1}. Thus, let us represent the equation \eqref{EqEinst0_1} in the form:
\begin{equation}\label{EqEinst0}
\dot{H}=-3H^2+\frac{e m^2\Phi^2}{2}-\frac{\alpha\Phi^4}{4}+\lambda.
\end{equation}

Introducing also the \emph{invariant cosmological acceleration}
\begin{equation}\label{omega}
\omega\equiv\frac{a\ddot{a}}{\dot{a}^2}\equiv 1+\frac{\dot{H}}{H^2},
\end{equation}
let us represent the equation \eqref{dH/dt} in the equivalent form:
\begin{equation}\label{omega<1}
\omega=1-\frac{1}{2}\frac{\dot{\Phi}^2}{H^2}\quad (\leqslant 1).
\end{equation}
Let us notice also useful relation for quadratic invariant of the Friedmann space's curvature \eqref{Freed}:
\begin{equation}\label{sigma}
\sigma\equiv \sqrt{R_{ijkl}R^{ijkl}}= H^2\sqrt{6(1+\omega^2)}\geqslant 0.
\end{equation}

\subsection{Standard Form of the Dynamic Equations}
Introducing dimensionless time variable $\tau =mt;\: \phi'=d\phi/d\tau$, dimensionless fundamental constants and dimensionless functions:
\begin{eqnarray}\label{rename}
\alpha_m=\frac{\alpha}{m^2}; \quad \lambda_m=\frac{\lambda}{m^2};\quad h=\frac{a'}{a}=\frac{H}{m},
\end{eqnarray}
let us rewrite the system of base equations \eqref{Eq_Phi0} and \eqref{EqEinst0} in standard form:
\begin{eqnarray}\label{Phi'}
\Phi'=Z\quad (=P_1);\\
\label{Z'}
Z'=-3hZ-e\Phi+\alpha_m\Phi^3\quad (=P_2);\\
\label{h'}
h'=-3h^2+\frac{e\Phi^2}{2}-\frac{\alpha_m\Phi^4}{4}+\lambda_m\quad (=P_3).
\end{eqnarray}
The expressions for invariant cosmological acceleration \eqref{omega} and \eqref{omega<1} remain invariant with the substitution $H\to h$ and derivatives' substitution $d/dt\to d/d\tau$, while the expression for the effective energy \eqref{E} is transformed in the following way:
\begin{eqnarray}\label{E_m}
\mathcal{E}=m^2 \mathcal{E}_m; \nonumber\\
\mathcal{E}_m(\Phi,Z)=\frac{Z^2}{2}+\frac{e\Phi^2}{2}-\frac{\alpha_m\Phi^4}{4}+\lambda_m\geqslant 0.
\end{eqnarray}
Let us write the integral condition \eqref{EqEinst0_4_E} in the same notation
\begin{eqnarray}\label{hE}
3h^2-\frac{Z^2}{2}-\frac{e\Phi^2}{2}+\frac{\alpha_m\Phi^4}{4}-\lambda_m=0.
\end{eqnarray}
Let us note, that in the notation taken all variables and constants are dimensionless.

Before we proceed to qualitative analysis of the dynamic system \eqref{Phi'} -- \eqref{h'} with integral condition \eqref{EqEinst0_4_E}, let us note the symmetrical properties of this system - namely its invariance with respect to transformations
\begin{eqnarray}\label{transform1}
\{\tau\to-\tau, & \displaystyle \Phi\to \Phi,\quad Z\to-Z,\quad h\to -h\};\\
\label{transform2}
\{\tau\to-\tau, & \displaystyle \Phi\to -\Phi,\quad Z\to Z,\quad h\to -h\};\\
\label{transform3}
\{\tau\to\tau, & \displaystyle \Phi\to -\Phi,\quad Z\to -Z,\quad h\to h\}.
\end{eqnarray}
These properties of symmetry allow simplification of the considered dynamic system's research.

\section{The Qualitative Analysis of the Dynamic System}
\subsection{Singular Points of the Dynamic System}
The considered dynamic system has 3 degrees of freedom and its condition is uniquely defined by coordinates of point $M(\Phi,Z,h)$ $\equiv M(x_1,x_2,x_3)$ in 3-dimensional phase space $\mathbb{R}^3$. Herewith phase space's areas where energetic condition \eqref{E_m} is violated, i.e., in which it is
\begin{equation}\label{E<0}
\Omega\subset \mathbb{R}^3:\quad \mathcal{E}_m(\Phi,Z)<0,
\end{equation}
are prohibited for the dynamic system. As a result, the phase space of the dynamic system can turn out to be multiply connected. Let us notice that prohibited areas \eqref{E<0}, if such do exist, are cylindrical ones with axes parallel to $Oh$. As a consequence of the Einstein equation \eqref{hE} such areas can exist \emph{only} in a plane $h=0:\ \{\Phi,Z\}$, i.e., only in points of the Universe's extension/compression stop. Then, the singular points of the dynamic system are defined by equality of right parts of the normal system of equations.

Thus, for the coordinates of singular points of the dynamic system \eqref{Phi'} -- \eqref{h'} we have the following equations:
\begin{eqnarray}\label{Eq_points}
Z=0;\;  -3hZ-e\Phi+\alpha_m\Phi^3=0;\\
 -3h^2+\frac{e\Phi^2}{2}-\frac{\alpha_m\Phi^4}{4}+\lambda_m=0.\nonumber
\end{eqnarray}
Thus, in a general case we have 6 singular points of the system -- two symmetrical ones with null potential and its derivative:
\begin{equation}\label{M_pm}
M_\pm\biggl(0,0,\pm\sqrt{\frac{\lambda_m}{3}}\biggr),\quad(\mathrm{if}\; \lambda_m\geqslant0, \forall\alpha)
\end{equation}
and 4 symmetrical $M_{ab}$, situating in vertices of rectangle\footnote{Let us agree on the notation: $M_{11}(-,0,-)$; $M_{12}(-,0,+)$; $M_{21}(+,0,-)$ and $M_{22}(+,0,+)$.} in plane $P_2(\Phi,h): Z=0$:
\begin{eqnarray}\label{M_11,12}
M_{ab}\biggl(\pm\frac{1}{\sqrt{e\alpha_m}},0, \pm \sqrt{\frac{\lambda_\alpha}{3}}\biggr),\nonumber\\
\mathrm{if}\;e\alpha>0;\; \lambda_\alpha\equiv \lambda_m+\frac{1}{4e\alpha_m}\geqslant0.
\end{eqnarray}
Let us notice that if we would use equation \eqref{dH/dt} instead  \eqref{h'}, the coordinate $h$ of singular points should have remained arbitrary which exactly is a consequence of the cited above arbitrariness.

Thus, in total, the following 5 cases are possible:\\
2.\; at $\{e\alpha>0,\lambda<-1/4e\alpha_m\}$ singular points do not exist;\\
3.\;  at $\{e\alpha<0,\lambda>0\}$ there exist only two singular points $M_\pm$;\\
4.\; at $\{e\alpha>0, -1/4e\alpha_m<\lambda<0\}$ there exist 4 singular points $M_{11},M_{12},M_{21},M_{22}$\\
5.\;  at $\{e\alpha>0,\lambda>0\}$ there exist all 6 singular points.

Calculating the value of effective energy \eqref{E_m} in singular points, we find:
\begin{equation}\label{E_m(M)}
\mathcal{E}_m(M_\pm)=\lambda_m;\quad \mathcal{E}_m(M_{ab})=\lambda_\alpha.
\end{equation}
Thus, from the conditions \eqref{M_pm} and \eqref{M_11,12} it follows: \emph{all singular points of dynamic system, if they exist, are accessible}.

\subsection{Character of Singular Points of Dynamic System}
Calculating the matrix of the dynamic system \eqref{Phi'} -- \eqref{h'}  $A_{ik}=\partial_kP_i$ where $P_i$ are right parts of the equations of the dynamic system, let us find with an account of \eqref{Eq_points}:
\begin{equation}\label{A}
A(M)=\left(\begin{array}{lll}
0                   & 1   &   0 \\
-e+3\alpha_m\Phi^2  & -3h &   0 \\
0                   &0    & -6h \\
\end{array}\right).
\end{equation}
The determinant of the matrix is equal to
\begin{equation}\label{detA}
\mathrm{det}(A)=-6h(e-3\alpha_m\Phi^2),
\end{equation}
and the characteristic equation for eigenvalues $k$ with respect to matrix \eqref{A} has the form:
\[-(k+6h)(k^2+3kh+e-3\alpha_m\Phi^2)=0.\]
Thus, with an account of integral condition \eqref{hE} the eigenvalues of the matrix are equal to:
\begin{eqnarray}\label{k_i}
k_1=-6h;\; k_{2,3}=-\frac{3}{2}h\pm \\
\frac{1}{2}\sqrt{\frac{3e\Phi^2}{4}+3\lambda_m+12\alpha_m\Phi^2-4e}.\nonumber
\end{eqnarray}
The eigenvalues of the matrix of the dynamic system in singular points are:
\begin{eqnarray}\label{eigen}\nonumber
\begin{array}{l}
M_\pm:\quad k_1=\mp  2\sqrt{3\lambda_m},\nonumber\\[6pt]
k_{2,3}=\mp\frac{1}{2}\sqrt{3\lambda_m}\pm \frac{1}{2}\sqrt{3\lambda_m-4e};\\[6pt]
M_{ab}:\quad k_1=\mp 2\sqrt{3\lambda_\alpha}; \\[6pt]
k_{2,3}=\mp\frac{1}{2}\sqrt{3\lambda_\alpha}\pm \frac{1}{2}\sqrt{3\lambda_\alpha+8e}.\\
\end{array}
\end{eqnarray}

The eigenvalues of matrix of the dynamic system in pairs of symmetrical points $(M_{11},M_{21})$ and $(M_{12},M_{22})$ coincide. The signs in front of first and second radicals in these formulas take values independent from each other: the signs in front of first radicals correspond to different pairs of points, signs in front of second radicals correspond to various eigenvalues \eqref{M_pm} and \eqref{M_11,12}.

Thus, singular points $M_\pm$ at $3\lambda_m-4e>0$ are instable (saddle) points and at $3\lambda_m-4e<0$ they are stable (attractive centers). The Hubble constant in these points is $h=\pm \sqrt{\frac{\lambda_m}{3}}$ at $\lambda>0$, scalar field is absent.
According to formulas \eqref{omega} and \eqref{sigma} the invariant cosmological acceleration and invariant curvature are equal to:
\begin{equation}\label{w,s_0}
\omega(M_\pm)=1;\quad \sigma(M_\pm)=\frac{2}{\sqrt{3}}\lambda.
\end{equation}
Thus, singular points $M_\pm$ are corresponded by inflationary solutions with positive or negative Hubble constant:
\[a\sim \mathrm{e}^{\pm \sqrt{\lambda/3}\ t}.\]
All eigenvalues of the dynamic system's matrix in its singular points $M_{ab}$ are real and have different signs at $e=+1$. Thus, all singular points at $e=+1$ are saddle. The eigenvalues of the dynamic matrix in its singular points $M_{ab}$ become complex conjugated at $e=-1$ and $\lambda_\alpha<8/3$ and there are values corresponding to attraction among them.

\section{Numerical Simulation}
\subsection{General Observations}
Let us proceed to numerical integration of the dynamic system \eqref{Phi'} -- \eqref{h'}. First of all, let us notice that this dynamic system is defined by enumerated list of three dimensionless parameters $\mathrm{P}=[e,\alpha_m,\lambda_m]$ and enumerated list of initial conditions
$I=[\Phi_0,Z_0,\epsilon]$, where $\epsilon=\pm 1$, and the value $\epsilon=+1$ is corresponded by the extension phase in the initial time instant $t_0$, while the value $\epsilon=-1$ is corresponded by the compression phase in this time instant. According to the above said, the initial value of the Hubble constant is defined by the Einstein equation \eqref{hE}, from which we have:
\begin{equation}\label{h0}
h_0=\frac{\epsilon}{3}\sqrt{\frac{Z^2_0}{2}+\frac{e\Phi^2_0}{2}-\frac{\alpha_m\Phi^4_0}{4}+\lambda_m}\equiv\frac{\epsilon}{3}\sqrt{\mathcal{E}^0_m}.
\end{equation}

Second, let us notice that dynamic system \eqref{Phi'} -- \eqref{h'} is an autonomous system of ordinary differential equations not depending explicitly on the time variable, therefore it is invariant with respect to translation $t\to t_0+t$. Therefore one can choose any value $t_0$ as an initial time instant when formulating initial conditions. We will lay it equal to zero. We are entitled to consider also the conditions of dynamic system at negative times $t_0<0$.
\subsection{Positive Values of the Cosmological Constant ($\lambda>0$)}
Let us consider in the beginning a simple case of existence of only two singular points $M_\pm$:
\begin{equation}\label{P1}
\mathrm{P}=[1,1,0.01]
\end{equation}
Herewith, let us establish initial conditions in the form $\mathrm{I}=[0.1,0.18,1]$.
\Fig{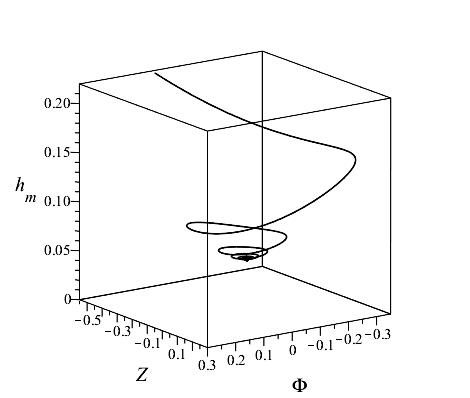}{8}{\label{Fig1}The phase trajectory of the dynamic system with parameters \eqref{P1} and initial conditions $\mathrm{I}=[0.1,0.18,1]$}
As it follows from the results of qualitative analysis, the dynamic system tends to the attracting non-zero focus $M_+$ along the compressing spiral.
The Universe asymptotically comes to constant extension velocity mode $h_\infty=\sqrt{\lambda_m/3}$, $\omega_\infty = 1$ (Fig. \ref{Fig1}).
Fig. \ref{Fig2} shows evolution of the cosmological acceleration for this case.
\Fig{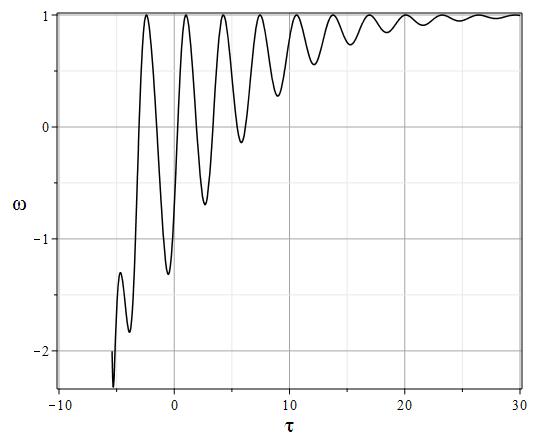}{6}{\label{Fig2}Evolution of the cosmological acceleration for the system with parameters \eqref{P1} and initial conditions $\mathrm{I}=[0.1,0.18,1]$}

However, dynamic system \eqref{Phi'} -- \eqref{h'} does not have the considered above simple behavior when Hubble constant falls with time to $+h_\infty$ in all cases. Fig. \ref{Fig3} shows the evolution of the dynamic system with positive parameters
\begin{equation}\label{P2}
\mathrm{P}=[1,1,0.1]
\end{equation}
and small initial values.
\Fig{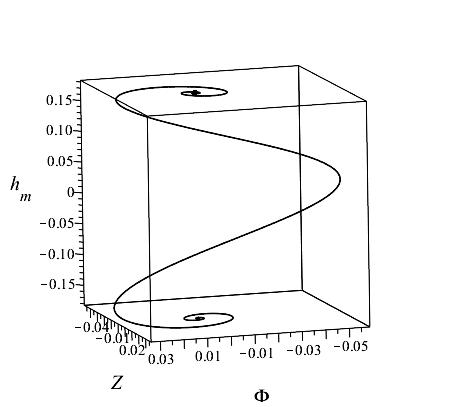}{9}{\label{Fig3}Phase trajectory of the dynamic system with parameters \eqref{P2} and initial conditions $\mathrm{I}=[0.00001,0.0001]$}
Fig. \ref{Fig4} shows the evolution of the Hubble constant for this case.
\Fig{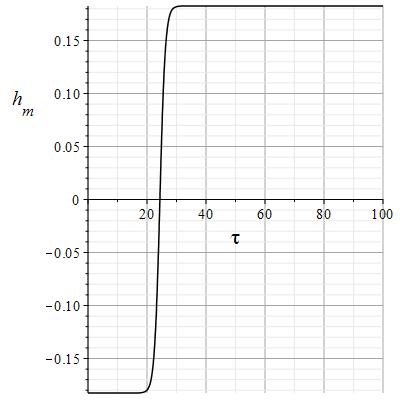}{9}{\label{Fig4}The evolution of the Hubble constant for the system with parameters \eqref{P2} and initial conditions $\mathrm{I}=[0.00001,0.0001]$}
Thus, the system starts from the neighbourhood of the singular point $M_-$ and then raises up to singular point $M_+$ over spiral line.

Let us now consider the case of positive parameters and large values of $\alpha$
\begin{equation}\label{P3}
\mathrm{P}=[1,20,0.01]
\end{equation}
In this case, there appear prohibited areas $\mathcal{E}_m(\Phi,Z)$ $<0 $ in a phase space and phase trajectories are being split by types of behavior (Fig. \ref{Fig5}) into 3 currents.

In the first current the phase trajectories press to the right boundary of the prohibited area. In the second current trajectories are winding onto attractive focus $M_+$. In the third current the trajectories press to the left boundary of the prohibited area. The phase trajectories of the first and third currents transition to the area of negative values of $h$, and trajectories of the third current tend to constant positive value $h(M_+)$ (Fig. \ref{Fig6}). Let us notice that namely the trajectories pressing to the boundary of prohibited area $\mathcal{E}_m=0$ do perform transition to the area of negative and unlimitedly growing in modulus values $h$.

\Fig{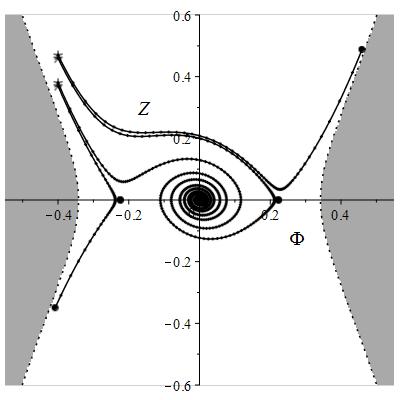}{8}{\label{Fig5}Phase trajectories of the dynamic system with parameters \eqref{P3} and initial conditions (top - down) $\mathrm{I}=[-0.4,0.47]$, $\mathrm{I}=[-0.4,0.46]$ and $\mathrm{I}=[-0.4,0.37]$. The plane $\{\Phi,Z\}$ is shown. The projections of 3 singular points are marked with black circles on the axis $\Phi$, while the central one of those is hidden by the trajectory.}
\Fig{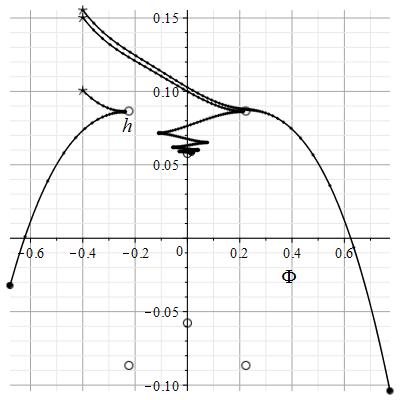}{8}{\label{Fig6}Phase trajectories of the dynamic system with parameters \eqref{P3} and initial conditions (top - down) $\mathrm{I}=[-0.4,0.47]$, $\mathrm{I}=[-0.4,0.46]$ and $\mathrm{I}=[-0.4,0.37]$. The plane $\{\Phi,h\}$ is shown. The singular points of the system are marked with empty circles.}
\subsection{Negative Values of the Cosmological Constant ($\lambda<0$)}
In this particular case singular points $M_\pm$ are absent. Lt us for the beginning consider the case clearly demonstrating the source of the mistake, which led Author to the statement about existence of Euclidian cycles (Fig. \ref{Fig7})
\begin{equation}\label{P4}
\mathrm{P}=[1,1,-0.01].
\end{equation}

It is seen from this graph that actually phase trajectory in the beginning is pressed significantly to the boundary of the prohibited area $\mathcal{E}_m=0$, which allows, at first glance, an interpretation of this process as a process of origination of the Euclidian cycle. However, the trajectory, unwinding, starts to leave the boundary of the prohibited area. More distinctively this process can be understood on the basis of 3-dimensional phase portrait (Fig. \eqref{Fig8}).

One can see on the Fig. \ref{Fig8} that when the pressing of the phase trajectory to the boundary of prohibited area is maximum
$\mathcal{E}_m=0$ the velocity of the extension $h$ turns to null.

\Fig{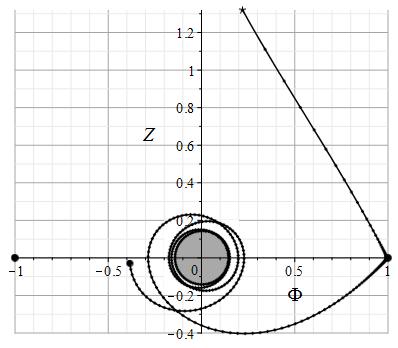}{7}{\label{Fig7}Phase trajectories of the dynamic system with parameters \eqref{P4} and initial conditions $\mathrm{I}=[0.945,0.1]$. The plane $\{\Phi,Z\}$ is shown. The projections of 2 singular points are marked with black circles on the axis $\Phi$. The initial point is marked with a star, the finishing point is marked with a black circle.}
\Fig{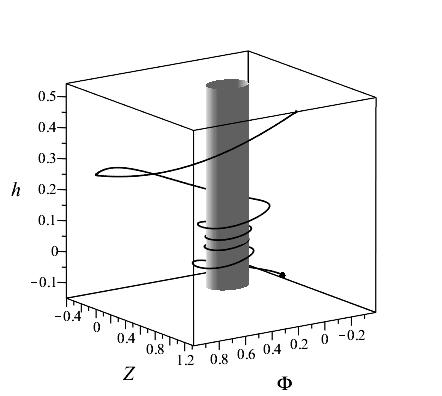}{7}{\label{Fig8}Phase trajectory of the dynamic system with parameters \eqref{P4} and initial conditions $\mathrm{I}=[0.945,0.1]$. The cylindric boundary of the prohibited area is depicted.}

However, the evolution $h(\tau)$ does not finish with this -- $h(\tau)$
continuously transitions into area of negative values corresponding to the compression phase. The condition $h(\tau)\geqslant 0$ that seemed to be obvious led the Author, as a result of numerical integration of the dynamic system, to the incorrect conclusion about infinite process of approaching of the phase trajectory to the boundary of area with null energy, i.e., to the conclusion on possibility of Euclidian cycles. Thus, the notice by Sergey Vernov turned to be perfectly correct.

\Fig{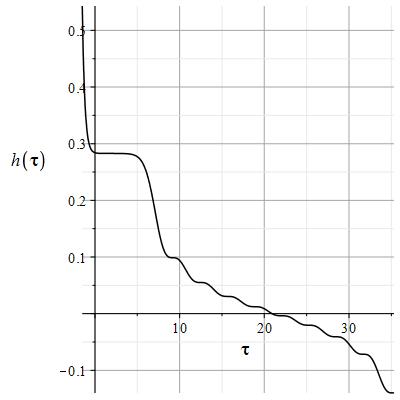}{8}{\label{Fig9}Evolution of the Hubble constant for the system with parameters \eqref{P4} and initial conditions $\mathrm{I}=[0.945,0.1]$}
The illustrated dependency $h(\tau)$ on Fig. \ref{Fig9} demonstrated this continuous transition as well as the fact that point $h(\tau_0)=0$ is a point of inflexion of the graph, which even more increased the likelihood of the conclusion about transition of the system to the regime of Euclidian cycle. Let us show that in plane $Z=0$ point $h(\tau_0)=0$, actually, is an extreme point. Tt follows from $h=0$  that $\mathcal{E}_m=0$, however then in plane $Z=0$ according to \eqref{E_m} it is
\[\frac{e\Phi^2}{2}-\frac{\alpha_m\Phi^4}{4}+\lambda_m= 0.\]
Then from the equation \eqref{h'} it follows $h'=0$, quod erat demonstrandum.

Let us consider the case of more complex topology of the system's space which appears, for instance, at large values $\alpha_m$;
\begin{equation}\label{P5}
\mathrm{P}=[1,20,-0.01].
\end{equation}
\Fig{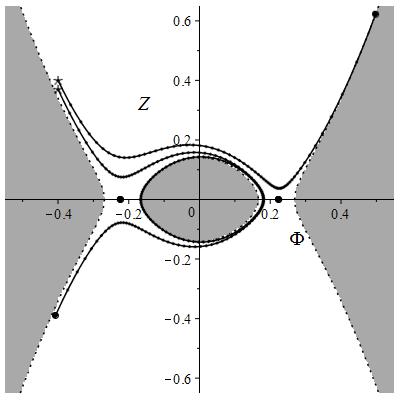}{9}{\label{Fig10}Phase trajectories of the dynamic system with parameters \eqref{P5} and initial conditions (top - down) $\mathrm{I}=[-0.4,0.47]$, $\mathrm{I}=[-0.4,0.37]$. The plane $\{\Phi,Z\}$ is shown. The second trajectory is being two times winded around central prohibited area.}
\Fig{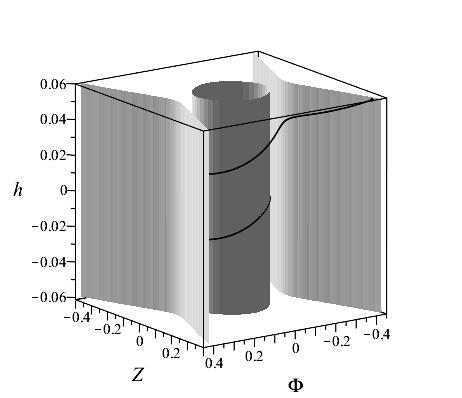}{7}{\label{Fig11}Phase trajectory of the dynamic system with parameters \eqref{P5} and initial conditions $\mathrm{I}=[-0.4,0.37]$ on the background of prohibited areas.}

The above cited Fig. \ref{Fig10} -- \ref{Fig11} also demonstrate the tendency of phase trajectories to adherence to the boundary of prohibited areas $\mathcal{E}_m=0$. More demonstrably this process can be shown on the graph of evolution of the effective energy $\mathcal{E}_m(\tau)$ (Fig. \ref{Fig12}).

\Fig{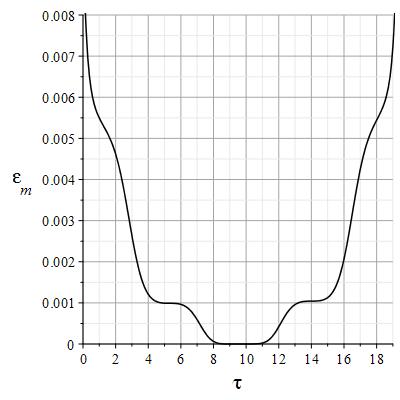}{7}{\label{Fig12}Evolution of effective energy $\mathcal{E}_m$ of the dynamic system with parameters \eqref{P5} and initial conditions $\mathrm{I}=[-0.4,0.37]$.}

\section{Discussion of the Results}
To summarize the results of this paper, let us notice the following, most important findings:
\textbf{1.} The research has shown that mathematical model of cosmological evolution of scalar fields with Higgs potentials, 
laid in the basis of the Author's and his students' researches \cite{IgnKokh1} -- \cite{IgnAgaf1} was mathematically 
incorrect in a part of equivalence of the considered dynamic system to the complete system of Einstein and scalar field equations. 
The assumption of non-negativity of the extension velocity $H(t)$ in certain cases contradicts to the initial system of equations. 
At the same time, the results obtained in the previous papers concerning phase trajectories that are significantly far from 
the boundaries of prohibited areas, are trustworthy.

\textbf{2.} It has been shown  on the basis of complete system of equations that in the case of positive values of 
cosmological constant there exist phase trajectories of the system's cosmological evolution which originate in the 
neighbourhood of one of two singular points with null values of scalar potential and its derivative, and finish in 
the second symmetrical singular point. The symmetrical properties of the dynamic system \eqref{transform1} -- \eqref{transform3} 
allow us to speak about reversibility of this process over time. Thus, in case of positive values of the cosmological constant, 
the inflationary compression can be the initial stage of cosmological evolution while the finishing one would be inflationary 
extension, and vice versa.

\textbf{3.} At positive values of the cosmological constant and sufficiently large values of the constant of interaction of the scalar field $\alpha$, the appearance of prohibited areas $\mathcal{E}<0$ in the phase space of the dynamic system becomes a significant factor of the cosmological evolution.
By virtue of the factor of multiple connectivity of phase space of the dynamic system, the phase trajectories are split into dynamic currents with various limiting properties. There are currents of trajectories adhering to the boundary of prohibited areas among such currents. Namely these trajectories correspond to a change of a sign of extension's velocity $H$, i.e. to a change of extension mode to compression, and vice versa.

\textbf{4.} At negative values of the cosmological constant the tendency of the extension velocity's  sign change in course of cosmological evolution takes universal character. Herewith all the main conformities of the system's motion in the neighbourhood of the prohibited areas conserve.
The cited research, first and foremost, show the necessity of revision of all author works dedicated to cosmological evolution of scalar fields with Higgs potentials, in particular, phantom fields and asymmetrical scalar doublets. Secondly, we have to clarify the impact of the new factor -- change of cosmological evolution mode  ``compression -- extension'' and ``extension -- compression'' on the cosmology of the early Universe.  Within this context, possibly, we would have to return to the idea of cyclic Universe.

\subsection*{Acknowledgment}

 Authors expresses his gratitude to Sergey Vernov who made the Author acquainted with these researches during the Winter School-Seminar ``Petrov School 2018'' dated November, 2018.

 \subsection*{Funding}

 This work was funded by the subsidy allocated to Kazan Federal University for the
 state assignment in the sphere of scientific activities.


%

\end{document}